\documentclass{aa}
\usepackage{txfonts}
\usepackage{natbib}
\usepackage{graphicx}
\usepackage{longtable}
\bibpunct{(}{)}{;}{a}{}{,}

\begin{document}

\title{Multiwavelength study of Cygnus A IV. Proper motion and location of the nucleus}

\author{K.C.\ Steenbrugge\inst{1,2}
 \and  K.M.\ Blundell\inst{2}
 \and  S. Pyrzas\inst{1}
}

\offprints{katrien.steenbrugge@gmail.com}

\institute{ 
	 Instituto de Astronom\'{i}a, Universidad Cat\'{o}lica del Norte, Avenida Angamos 0610, Antofagasta, Chile 
	 \and
         Department of Physics, University of Oxford, Keble Road, Oxford OX1 3RH, UK}

\date{\today}

\abstract
{Cygnus A, as the nearest powerful FR II radio galaxy, plays an important role in understanding jets and their impact on the surrounding intracluster medium.}
{To explain why the nucleus is observed superposed onto the eastern lobe rather than in between the two lobes, and why the jet and counterjet are non-colinear. }
{We made a comparative study of the radio images at different frequencies of Cygnus A, in combination with the published results on the radial velocities in the Cygnus A cluster.}
{From the morphology of the inner lobes we conclude that the lobes are not interacting with one another, but are well separated, even at low radio frequencies. We explain the location of the nucleus as the result of the proper motion of the galaxy through the cluster. The required proper motion is of the same order of magnitude as the radial velocity offset of Cygnus~A with the sub-cluster it belongs to. The proper motion of the galaxy through the cluster likely also explains the non-co-linearity of the jet and counterjet.}
{}

\keywords {galaxies:active--galaxies:individual: Cygnus~A--galaxies:jets}

\titlerunning{}
\authorrunning{K. C. Steenbrugge et al.}

\maketitle

\section{Introduction \label{sect:intro}}

Cygnus~A (3C~405) has been well studied at radio wavelengths as it was one of the first double sources detected at radio wavelengths (see for instance \citealt{carilli91} and \citealt{carilli96}). At GHz frequencies and above a jet directed at $\sim$60$^{\circ}$ to our line of sight, embedded within the western lobe, is observed. A weaker counterjet approximately anti-parallel to the jet is observed within the counterlobe or eastern lobe. In a set of previous papers we have studied the jets and lobes of Cygnus~A using radio and X-ray images and X-ray spectra. We showed that the jets are precessing, have a best fit speed of 0.35~c, and that the angle between the jet and inner counterjet is 179$^{\circ}$, determined from the best-fit precession model \citep{steenbrugge08a}. From a visual inspection of the jet and counterjet, these authors determined a somewhat smaller, angle of 177$^{\circ}$$\pm$1$^{\circ}$.5. This result is unexpected and remains unexplained. 40~kpc out from the nucleus the counterjet bends through a 27$^{\circ}$ 31$^{\prime}$ angle, the cause of which is still poorly understood, but is possibly related to the presence of relic plasma \citep{steenbrugge10}. 

We also found that the plasma from a previous episode of jet activity explains the presence of a relic counterjet in the X-ray image of Cygnus~A \citep{steenbrugge08b} and the fact that at low frequencies the counterlobe is {\it brighter} than the lobe \citep{steenbrugge10}. Although a well known general result, important for this article is that the size of the lobe and counterlobe do depend on the observed frequency. Both the lobe and counterlobe are much larger at 151~MHz than at 5~GHz and are larger at 5~GHz than at 15~GHz (see for instance figure~5 in \citealt{steenbrugge10}). 
%Finally, we have shown that the lobes observed at 151 and 327~MHz do not have the pressure to clear out the intracluster medium, but that the remaining cluster gas is cleared out by the lobe as observed at $\sim$5~GHz \citep{steenbrugge11}. 

This radio galaxy is embedded in the Cygnus cluster formed by the merging of 2 clusters, each probably having a richness class 1 \citep{markevitch99,ledlow05}. The ongoing merger is clearly observed in the X-ray temperature map and supported by the radial velocity clustering and location of the galaxies \citep{markevitch99,belsole07,ledlow05}. From the radial velocity structure, the location of the galaxies, and the observed X-ray temperature structure, \cite{ledlow05} estimate that core encounter will occur in 0.2$-$0.6~Gyr hence. The average radial velocity of the cluster galaxies identified by \cite{ledlow05} is 19,008$\pm$160~km~s$^{-1}$ with a velocity dispersion of 1126~km~s$^{-1}$. For Cygnus A these authors measured a radial velocity of 16,811~km~s$^{-1}$, and thus an offset of 2197~km~s$^{-1}$ from the cluster average. However, the measured radial velocities are clustered in two groups, one with an average radial velocity of 17,648~km~s$^{-1}$ and one with an average radial velocity of 20,126~km~s$^{-1}$ (as determined by \citealt{ledlow05} using location and radial velocity information). From its radial velocity, the host of this powerful FR~II classical double radio galaxy is at a redshift of 0.05607 \citep{owen97}. Assuming a Hubble constant of 73 km s$^{-1}$ Mpc$^{-1}$, 1$^{\prime\prime}$ corresponds to 1.044~kpc.

Cygnus~A is not in the centre of either of the subclusters \citep{owen97}, and actually shows an offset of 72 $h^{-1}_{75}$~kpc from the centre of the smaller subcluster \citep{ledlow05}. The centres of the two subclusters are separated by 458~$h^{-1}_{75}$~kpc \citep{ledlow05}.
%According to these authors this is also the subcluster for which the radial velocity distribution in not well fit by a Gaussian.   
There is an overall temperature difference discerned in the cluster, due to the merger. At the location of Cygnus~A, but mostly on larger scales than the radio lobes, an asymmetry in temperature is clearly observed \citep{belsole07}. Due to the difference in timescale between the cluster merger and the jet age of the current activity, and the fact that the asymmetry in the intracluster medium (ICM) is on larger scales, the radio lobes are not affected by the ongoing merger process.  

In this paper we study the morphology of the region surrounding the nucleus using radio images at different frequencies. In aligning the radio and X-ray data, it is clear that the nucleus of the Cygnus~A galaxy does not lie between both lobes, but actually lies in front of, within or behind the counterlobe. Explaining this result, we find a likely explanation for a second observational result, namely that the jet and inner counterjet are not co-linear \citep{steenbrugge08a}.  

In the following Section we summarize the data used for this analysis. In Section~3 we describe the morphology of the inner lobes and the nucleus. The discussion is given in Section~4 and our conclusions are stated in Section~5.

\section{Data cleaning}

For this study we use the same radio data we used in previous papers by our group. The data are described in \cite{steenbrugge10}. A detailed description of the data processing can be found in \cite{steenbrugge08a} and \cite{steenbrugge10}. To summarise, we use the 151-MHz Multi-Element Radio-Linked Interferometer Network (MERLIN) image, which was kindly provided by P. Leahy and published by \cite{leahy89}. The 327~MHz, 1345~MHz, 5~GHz, 8~GHz and 15~GHz VLA\footnote{ The Very Large Array is a facility of the National Radio Astronomy Observatory, National Science Foundation.} data were obtained by C. Carilli and published by \cite{carilli91,carilli96}, \cite{carilli96b} and \cite{perley96}. Data from the four VLA configurations were combined to obtain the highest resolution possible without losing the extended emission.

\section{Morphology of the nuclear region}

\begin{figure}
\begin{center}
 \resizebox{\hsize}{!}{\includegraphics[angle=0]{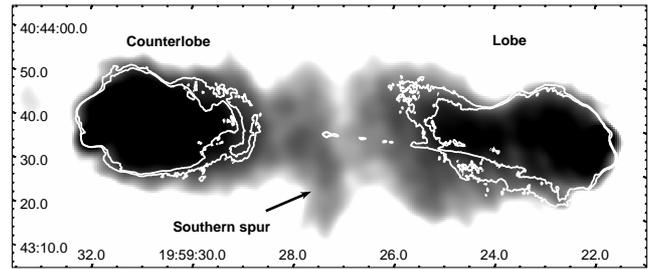}}
 \caption{151~MHz image showing the low surface brightness emission from the lobes as a grey-scale image. The white contours are 2 low surface brightness (0.0037 and 0.01 Jy/beam) levels from the 5~GHz image.  \label{fig:151_5GHz_tot}}
\end{center}
\end{figure}

\begin{figure}
\begin{center}
\resizebox{\hsize}{!}{\includegraphics[angle=0]{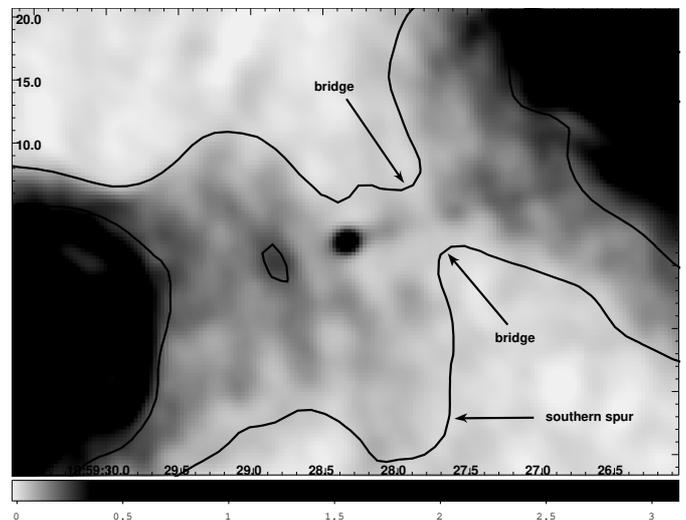}}
\caption{1345~MHz image of the nucleus overlaid with the 327~MHz
  contours between 2.5 and 20~Jy. Note that there is a ``bridge'' of
  emission between the lobe and counterlobe, where the width of the
  327~MHz emission is smallest. The Grey-scale indicates the intensity
  in Jy/beam. \label{fig:nucleus2}}
\end{center}
\end{figure}

To study the nuclear region in detail, we made several figures overlaying low- and high-frequency radio images. At low frequencies, 151 and 327~MHz, the nucleus is not detected consistent with the spectral shape of radio nuclei, e.g.\ \cite{pearson92}. On the other hand, emission from the lobes is detected near the location of the nucleus. At 1.345~GHz the nucleus is detected as well as some weak emission from the lobes near the nucleus. At frequencies higher than 1.345~GHz the nucleus and the jet are easily observed, but the counterjet is only weakly detected. However, at these frequencies no emission from the lobes is detected near the nucleus. 
%This is consistent with the aging of the synchrotron emitting electrons. The highest energy electrons are close to the hotspots, where the electrons are re-accelerated before entering the lobes. 
Figure 5 of \cite{steenbrugge10} clearly shows the difference in lobe size with radio frequency. In most figures in this article we use the 151~MHz and 5~GHz data, as these are the best quality images at low and high frequencies available. 

In Fig.~\ref{fig:151_5GHz_tot} we show in grey-scale the 151~MHz emission from the lobes, with a transfer function that shows the low surface brightness emission. Overlayed are two low-surface brightness contours of the 5~GHz image of Cygnus~A. The nucleus and part of the inner jet are detailed by the contours. From the figure it is clear that the counterlobe and lobe are well separated. However, contrary to expectations the nucleus is not located equidistantly from both lobes. Rather, the nucleus lies at the edge of, or in the counterlobe. From the 5~GHz image the asymmetry in distance between both lobes seems smaller.

\begin{figure}
\begin{center}
  \resizebox{\hsize}{!}{\includegraphics[angle=0]{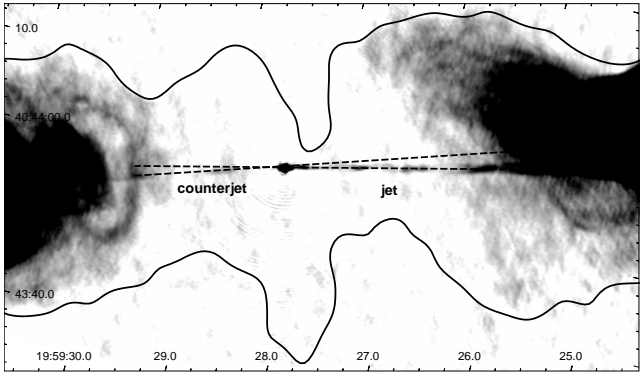}}
\caption{Detail of the nucleus and inner jets as observed at 5~GHz with
  a 151~MHz low brightness contour overlaid. The contour is at 4.5 Jy/beam. The dashed lines trace the jet (counterjet) extended to the counterlobe (lobe) side. The first weak jetknot in the counterlobe lies below the jet extended into the counterlobe. \label{fig:nucleus3}}
\end{center}
\end{figure}

To study the location of the nucleus with respect to the counterlobe we made two more figures, both detailing the nucleus and its surroundings only. In Fig.~\ref{fig:nucleus2} we show the 1345~MHz image of the nucleus and its surroundings as a grey-scale image with low surface brightness 327~MHz contours overlayed. From Fig.~\ref{fig:nucleus2} it is clear that the nucleus and thus the host galaxy, defined as the gravitationally bound structure, does lie just within the counterlobe, and again confirms that the nucleus certainly is not located equidistantly from both lobes. Defining the exact boundary of the lobes is somewhat complicated by the fact that there is very weak emission at 1.345~GHz, 327 and 151~MHz between both lobes, which we will call ``bridge''. This weak emission is narrower than the lobe emission further out (see Fig.~\ref{fig:nucleus2}) and is therefore likely related to the jet. An additional complication is that the resolution at 151 and 327~MHz is poorer than at 1345~MHz, 5 or 8~GHz. To further ascertain the position of the nucleus we also overlayed a low brightness 151~MHz contour on the 5~GHz image, which is shown in Fig.~\ref{fig:nucleus3}. Of the 3 figures this one most clearly shows that the nucleus is located within the low frequency emitting counterlobe. 

To gain a more quantitative idea of the difference in distance between the nucleus and the inner lobes, we measured the distance between the nucleus and the lowest radio contour for the lobes for 1.345~GHz, 5~GHz and 8~GHz using the jet to determine the direction towards both lobes. This we can compare to the distance between the nucleus and the inner edge of each lobe as measured at 151~MHz. For simplicity we assume that the jet and counterjet make a 180$^{\circ}$ angle. The inner edges of the lobes are not as sharp and well defined as the lobe edges near the hotspots, creating an uncertainty as to where the lobe emission reaches at the different frequencies. This is the largest uncertainty in the determined distances. We should note that the distance between the nucleus and the inner lobes varies with angle in the plane of the sky assumed for the measurement. Taking the jet direction allows us to easily compare the results at the different frequencies, but for some frequencies, this is not the smallest distance between the nucleus and the lobe. The distances between the nucleus and the inner edges of the lobes is dependent on the quality of the data, however, the difference in distance should be less affected by this. Therefore we only list the difference between both distances in Table~\ref{tab:distances}, and use these values in our comparison.

\begin{table}
\begin{center}
\caption{The difference in distance between the nucleus and the inner edge of the counterlobe and of the lobe along the jet direction for different radio frequencies. The difference in distances is given in observed arcseconds as well as a physical distance assuming that the jet and counterjet make a 60$^{\circ}$ angle in the plane of the sky.}
\label{tab:distances}
\begin{tabular}{lll}
\hline
\hline
frequency  & d (in $^{\prime\prime}$) & d (in kpc) \\\hline
151~MHz    & $>$7.9                 & $>$16.5 \\
1.345~GHz  & 5.8$\pm$1.7            & 11.6$\pm$3.5 \\
5~GHz      & 6.2$\pm$0.8            & 12.9$\pm$1.7 \\
8~GHz      & 6.6$\pm$2.3            & 14.0$\pm$4.9 \\
\hline
\end{tabular}\\
\end{center}
\end{table} 

At 1.345~GHz the distance between the inner edge of the lobe and the nucleus is 10$^{\prime\prime}$.3$\pm$1$^{\prime\prime}$.2. The distance between the nucleus and the inner edge of the counterlobe is 4$^{\prime\prime}$.5$\pm$0.$^{\prime\prime}$5. Taking a line-of-sight angle of 60$^{\circ}$ \citep{steenbrugge08b}, we obtain physical distances of 21$\pm$2.5~kpc and 9.4$\pm$1.0~kpc, and a difference in these distances of 11.6~kpc.  The results of the difference in distances are summarised in Table~\ref{tab:distances}. At 5~GHz the distance between the edge of the counterlobe and nucleus is 13$^{\prime\prime}$.6$\pm$0.$^{\prime\prime}$3, while the distance between the nucleus and the edge of the lobe is 19$^{\prime\prime}$.8$\pm$0$^{\prime\prime}$.5. This gives physical distances of 28.4$\pm$0.6 and 41.3$\pm$1.0~kpc. Thus even at 5~GHz the nucleus is not equidistant from the inner edge of the lobe, and the difference between both distances is a considerable 12.9~kpc.

At higher frequencies the lobes show more substructure, thus the chosen angle in the plane of the sky, i.e. along the jet direction and its extension on the counterlobe side, has a larger influence on the measured distances. At 8~GHz for the lobe side the distinction between where there is only jet emission and where the lobe plasma starts is somewhat ambiguous, partially because this is also a noisier image. For the 8~GHz image we find distances from the nucleus to the inner counterlobe and to the inner lobe of respectively 20$^{\prime\prime}$.4$\pm$1$^{\prime\prime}$ and 27$^{\prime\prime}$.0$\pm$1$^{\prime\prime}$.3 or 42.4$\pm$2.1~kpc and 56.4$\pm$2.7~kpc. The difference between both distances is somewhat larger, 14~kpc, than the result for 5~GHz, but also has a larger uncertainty. Considering the uncertainties, these differences in distance are consistent with each other.  At 15~GHz the counterlobe emission lies almost completely to the north of the extended jet line, and completely to the north of the current counterjet and 2 hotspots, due to the bending of the counterjet about midway to the hotspots. If we still measure distances to the inner edge of the lobes along the extended jet line, we find 50$^{\prime\prime}$.9 and 59$^{\prime\prime}$.6 or 106.3 and 124.4~kpc, and thus a difference of 18.1~kpc. Taking the closest emission to the nucleus for the counterlobe the distance decreases to 39$^{\prime\prime}$.7 or 82.9~kpc, which is more than 20~kpc closer than along the extended jet line. If on the other hand we follow the current counterjet, then because the emission from the lobe lies completely north of the counterjet trajectory, we only encounter emission from the smaller hotspot. Considering these large differences we do not give a value for the difference in distance for this frequency in Table~\ref{tab:distances}. 

At 151 and 327~MHz the nucleus and jet are not detected. To obtain the distance between the nucleus and the inner edge of the lobe we use the position for the nucleus and the direction of the jet as determined from the 5 GHz image. The distance between the nucleus and the inner edge of the lobe is 7$^{\prime\prime}$.9 or 16.5~kpc, but again there is a rather large uncertainty. Assuming a distance of 0~kpc between the nucleus and the inner edge of the counterlobe, the above distance is also the difference in distance. From Fig.~\ref{fig:nucleus3} it appears that actually the nucleus is located a few kpc within the counterlobe, and thus the above distance is a lower limit. 
%However, due to the existence of the ``bridge'', determining this distance is not really possible with the current data. The nucleus is thus roughly $\sim$15~kpc closer to the counterlobe than to the lobe at the studied frequencies. 
The difference in distance is either constant or increases for higher frequencies, if we exclude the 151~MHz measurement. In any case, at all frequencies studied, the inner lobe is further from the nucleus than the inner counterlobe. 
%and this result is nearly independent or independent of the frequency in which which observe Cygnus~A.   

Figs.~\ref{fig:151_5GHz_tot},~\ref{fig:nucleus2} and~\ref{fig:nucleus3} all show that both lobes are well separated, even at the lowest frequency. There is no evidence that the lobes are colliding and/or interacting with each other. Any interaction would likely lead to a shock and potentially an increased magnetic field, and hence an increase in higher-frequency emission near the nucleus, which is not observed. At frequencies above 1.345~GHz there is no emission between both lobes, with the exception of emission that can be clearly identified as coming from the jet. At 151~MHz and 327~MHz there is very weak emission between both lobes identified in Fig.~\ref{fig:nucleus2} and labelled as ``bridge''. This weak emission has a width of 9$^{\prime\prime}$.6, and is thus resolved in the 151~MHz and 327~MHz images. Although too wide for being emission directly from the jet as observed at 5~GHz, it is likely the emission is related to the jet due to its morphology: narrower than either the lobe or counterlobe, and coinciding with the jet trajectory as detected at 1.345 and 5~GHz. Higher resolution and deeper low-frequency images are needed to clarify the origin of this emission.

\begin{figure}
\begin{center}
  \resizebox{\hsize}{!}{\includegraphics[angle=0]{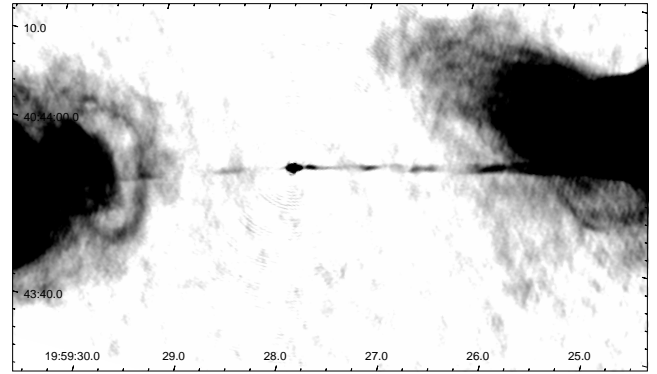}}
\caption{Detail of the nucleus and inner jet and counterjet as observed at 5~GHz, optimised for the detection of the knots in the counterjet.  \label{fig:nucleus4}}
\end{center}
\end{figure}

In Fig.\ref{fig:nucleus3} we draw with dashed lines the best fit straight line of the jet between the jet entering the lobe as detected at 5~GHz and the nucleus, and extending this line to the counterlobe side. We also draw a dashed line connecting the nucleus with the weaker counterjet knots reaching a weak ring-like feature in the 5~GHz (see Fig.~\ref{fig:nucleus3}) image and extending it to the lobe side. Consistent with the results of the precession modelling of the jet in \cite{steenbrugge08a}, the jet, nucleus and counterjet knots are not precessing around a straight line. 
%Considering the jets show curvature due to precession, the the difference in angle between both jets should really be determined from the precession model. 
From precession, the difference from a 180$^{\circ}$ angle is 1$^{\circ}$ \citep{steenbrugge08a} and thus somewhat smaller than the angle between both dashed lines. The dashed line overlaying the counterjet knots makes these very hard to see, therefore in Fig.~\ref{fig:nucleus4} we show the 5 GHz image without lines. 

%\begin{figure}
%\begin{center}
%\resizebox{\hsize}{!}{\includegraphics[angle=0]{5_327overlay_bw.ps}}
%\caption{Detail of the nucleus and inner jet as observed at 5~GHz with
%  the 327~MHz contours overlaid. The contours are at 2.5, 7 and 20
%  Jy/beam. The dashed line traces the jet extended to the counterlobe side, showing that the first weak jetknot in the counterlobe lies below this extension. The grey-scale indicates the intensity in Jy/beam. \label{fig:nucleus5}}
%\end{center}
%\end{figure}

\section{Discussion}
In this Section we discuss possible explanations for the location of the nucleus, and thus also the gravitationally bound host the galaxy. The nucleus is superposed onto the counterlobe, indicating that it is located in front of, within or behind the counterlobe, rather than midway between the lobe and the counterlobe. We also discuss a possible explanation for the observed non-colinearity between the inner jet, nucleus and inner counterjet, as determined from the precession modelling \citep{steenbrugge08a}. 

\subsection{Why is there a nuclear position offset}
\subsubsection{Asymmetric hotspot advance speed and/or expansion of the lobe and counterlobe}
A possible explanation for the position of the nucleus is that the hotspot advance speed is different for the lobe and counterlobe, due to a difference in environment, i.e. the density and/or temperature of the ICM, into which the lobes expand. If the hotspot advance speed in the counterlobe is slower, then the larger density in the counterlobe will cause an increased expansion either sideways or toward the nucleus or both.  As the distance between the nucleus and the hotspot in the counterlobe is expanding more slowly, this also means that the distance the electrons in the counterlobe plasma need to travel to reach the nucleus is smaller. 
%In this scenario the counterlobe would be smaller, taking light-travel effects into account, and equally bright or brighter than the lobe at radio frequencies. The X-ray emission from the ICM on the counterlobe side would be brighter, have a higher temperature, or both than on the lobe side, due to the higher temperature/density of its ICM. 

To test this hypothesis we use the luminosities and volumes of both lobes at the different frequencies detailed in \cite{steenbrugge10}. At frequencies below GHz the counterlobe is indeed brighter, but at higher frequencies both lobes are equally bright. The counterlobe is smaller at frequencies below 8.5 GHz, but both lobes have the same size at 8.5~GHz and the counterlobe is larger at 15~GHz. Further, \cite{steenbrugge10} rule out that a difference in adiabatic expansion for both lobes can explain the excess low-frequency emission in the counterlobe. 
%The radio data thus gives a rather mixed picture. 

From large scale X-ray images we know that the Cygnus cluster is undergoing a merger \citep{markevitch99}. This merger should have increased the density on the lobe side of the cluster, and certainly has resulted in a higher temperature of the ICM on the lobe side \citep{belsole07}. Thus, if there is an effect on the hotspot advance speed due to differences in the ICM properties, it should be the lobe side hotspot that is slowed down. From the X-ray data there is no evidence that the advance speed of the hotspots in the counterlobe is slower than the advance speed of the hotspots in the lobe. Hence, we think it unlikely that any asymmetry in the hotspot advance speed could explain why the nucleus lies in the counterlobe.
%Roland, Pelletier and Muxlow 1988 and Muxlow 1988

A related possibility is that the hotspot advance speed is the same for both lobes, but that the expansion of the lobe is different due to density asymmetries. In this scenario, the counterlobe would have had an excess expansion toward the nucleus and a smaller expansion perpendicular to the jet axis. This would result in a narrower counterlobe in comparison with the lobe, assuming that the width of the lobe is the same as the depth of the lobe. The width of both lobes is rather variable, but the counterlobe is not obviously narrower. Furthermore, due to the merger, if there is a higher ICM pressure, we expect it to be on the lobe side. So we do not prefer this explanation either.

\subsubsection{Emission from the bridge and lobe-counterlobe interaction}
An alternative explanation might be that the nucleus is located inside both lobes, which have interacted in the past or are currently interacting. In this scenario we expect a brightening of the low-frequency radio emission, due to the increase in the electron density at the site of the interaction. We also expect a re-brightening of higher radio frequency emission, resulting from the compression of the plasma, re-energising the electrons. A likely increase in magnetic field strength would ensure that the electrons also emit synchrotron radiation at higher frequencies. Finally, if the interaction is supersonic, then the shocks should cause an increase in the X-ray determined temperature near the nucleus.

In high-frequency images the lobes are well separated with a clear lack of emission between them. At frequencies below GHz there is a narrow strip of very weak radio emission between both lobes, which we label the ``bridge'' in Fig.~\ref{fig:nucleus2}. This is the most likely location of any interaction between both lobes. However, there is certainly no brightening observed at low frequencies, on the contrary the ``bridge'' has the lowest flux density. Further morphological evidence against any interaction between both lobes comes from the fact that at 151~MHz the ``bridge'' is narrower than any part of the lobes further out from the nucleus. This is contrary to expectations if this is the site of lobe interaction.
%as then one would expect the width of the emission to be at least as wide as the width of the rest of the lobe. 
Moreover, the temperature of the X-ray plasma near the nucleus is the lowest of all measured temperatures in this cluster and these regions were described as ``belts'' by \cite{wilson06}. There is thus no evidence of shocks near the nucleus. We conclude that the counterlobe and lobe do not (yet) interact with each other.

The jet observed at higher frequencies passes through the ``bridge'' observed at 151 and 327~MHz (see Fig.~\ref{fig:nucleus2}), suggesting that the emission is related to the jet. However the width of the bridge, 9.6$^{\prime\prime}$ in both the 151 and 327~MHz images, cannot be explained by the width of the current jet as observed at 5~GHz and the lower resolution at 151 and 327~MHz. Higher resolution and deeper low frequency images are needed to determine the exact origin of this emission.
% but it could be adiabatically expanded emission from the region that is perturbed by the jet. The nucleus is not detected at 327 or 151~MHz .

%The most likely explanation for the ``bridge'' is that the emission is indeed from the jet, or due to the jet interacting with the ICM. The location of the nucleus in relation to the lobes is best seen in the 1345~MHz and 5~GHz images with the 327~MHz lobe low brightness contours overlaid as shown in Figs.~\ref{fig:nucleus2} and~\ref{fig:nucleus5}. As mentioned above, it seems that the nucleus is located within the counterlobe instead of being located as expected between both lobes. The ``bridge'' is the narrowest part of either lobe and lies to the west of the nucleus. The western lobe does seem to have an inner
%edge which is connected to the nucleus via the ``bridge'', and thus
%shows no evidence of lobe plasma being displaced. The eastern lobe
%however seems to have a southern extension observed only at low
%frequencies, which we will call the ``southern spur'' (see Fig.~\ref{fig:nucleus2}). Considering the
%narrowness of the ``bridge'' we think it highly unlikely that this
%``southern spur'' results from both lobes interacting. The
%``southern spur'' is discussed in section~\ref{sect:spur}.

\subsubsection{Buoyantly rising lobe}
The difference in distance between the nucleus and the inner edges of the lobes could be due to the fact that the current lobe has buoyantly risen, while the counterlobe has not. It seems unlikely that the lobe has risen by about 16.5~kpc, while the counterlobe has not risen at all. Furthermore, although this could in principle explain the asymmetrical distance between the nucleus and the inner edge of the lobe and counterlobe, it cannot explain why the nucleus is superposed onto the counterlobe, nor why the outer hotspot in the lobe is further from the nucleus than the outer hotspot in the counterlobe. 

\subsubsection{Projection effect}
We do not know whether the nucleus is located in the counterlobe or in front of it, and thus projected onto the counterlobe.  Depending on the depth of the counterlobe it could appear that the counterlobe engulfs the nucleus, while the nucleus in reality is not located in the counterlobe. For the nucleus to be equidistant from both lobes then, the counterlobe must have expanded in depth more than the lobe, at least near the nucleus. 

We cannot measure the depth of the lobe, therefore we assume that the depth of the lobe equals its width. The areas measured from low-frequency radio data, and the calculated volumes, are larger for the lobe than for the counterlobe \citep{steenbrugge10}. Assuming that the width of the lobe is also its depth, the lobe volume is calculated to be 1.12 times larger than the counterlobe at 151~MHz and 1.37 at 327~MHz \citep{steenbrugge10}. Unless the ICM density has an angular dependence, such that the expansion in the width direction of the lobe is different from the expansion in the depth direction, the above assumption should be valid on average. However, the variable width measured along the length of the lobes, is a clear indication of local differences in the ICM that can influence lobe expansion. On average the width of the lobe is somewhat larger than that of the counterlobe, and therefore it seems unlikely that the depth near the nucleus is significantly larger for the counterlobe than for the lobe.

If the counterlobe is longer, then even for the same width there could be a projection effect. One can measure the distance between the nucleus and the outer hotspots, which is independent of the frequency used. This distance is significantly longer for the lobe than for the counterlobe. The distance between the nucleus and the outer hotspot of the lobe and counterlobe, using the 5~GHz image, are 67$^{\prime\prime}$.3 and  58$^{\prime\prime}$.6. This gives a physical distance, assuming a line of sight of 60$^{\circ}$, of 140.6 and 122.3~kpc for the lobe and counterlobe respectively. Assuming a hotspot advance speed of 0.02c \citep{muxlow88} and using an angle of 60$^{\circ}$ between the jet axis and our line of sight, the light-travel time difference between the extreme hotspots is 2$\times$ 10$^5$ years \citep{steenbrugge08b}. We thus calculate that the length difference due to light travel time effects should be 1.2~kpc. For a hotspot advance speed of 0.005c \citep{alexander96} this value is 0.3~kpc. Both these measured hotspot advance speeds are much smaller than the average hotspot advange speed of 0.2 c found by \cite{best95}. However, \cite{best95} do note a large spread in the individual hotspot advance speeds, and thus this result is not inconsistent with the above measured hotspot advance speeds. Both values are much smaller than the actual measured difference in length of 18.3~kpc. This difference in length is of the same order, but larger, than the differences in distances between the nucleus and the inner edges of the lobes (see Table~\ref{tab:distances}). 
% Assuming that the lobes are in the plane of the sky halves the distances and the difference between both distances, but then there would be no difference in length due to the light travel time toward us. So the lobe is still significantly longer than the counterlobe. 

We have assumed above that the length is given by the distance between the nucleus and the furthest hotspot.  Generally, lobe emission does not fill the whole area between the nucleus and the hotspots, except for low power radio galaxies \citep{blundell99}. For the counterlobe as observed at 151~MHz the distance between the nucleus and the furthest hotspot using the 5~GHz image is also the length of the counterlobe. However, the lobe emission comes from an area with a length smaller than the distance between the nucleus and the outer hotspot at all frequencies. The length of the lobe at 5~GHz between the innermost emission, which is located away from the jet, and the furthest hotspot is 60.4$^{\prime\prime}$ or 126.1~kpc. If we measure the length of the lobe at 151~MHz, this time along the jet axis, we obtain a size of 60.3$^{\prime\prime}$ or 125.9~kpc. Both these values are still larger than the measured counterlobe length (122.3~kpc), but the difference in size is much smaller, and is only three times larger than the expected difference in length due to finite velocity of light. 
%The 9.3~kpc difference in length minus the difference due to light travel time divided by two is rather similar to the 5~kpc distance that we assume that the nucleus has travelled below. 
We conclude that the required excess counterlobe expansion toward the nucleus is unlikely, considering that the lobe is longer. Therefore projection effects alone cannot explain the fact that the nucleus appears to lie within the counterlobe.

\subsubsection{Proper motion}
A final explanation for why the nucleus is observed superposed onto the counterlobe is that the host galaxy and the nucleus have a proper motion through the cluster and sub-cluster. Because the hotspots and lobes are not gravitationally bound, they will not share this motion.  Thus, the nucleus could have moved into or in front of the counterlobe, which is presumed to be stationary with respect to the ICM. To test this scenario we will assume that originally the nucleus was located midway between the lobe and counterlobe, and was not engulfed by either. If we suppose that the nucleus moves along the plane of the sky toward the direction of the counterlobe, then it would have travelled $\sim$5 kpc for the nucleus to be located in front of the counterlobe. Considering that the measured radial velocity offset is toward us \citep{ledlow05}, it is likely that the nucleus is located in front of the counterlobe, and there is a projection effect in addition to the proper motion of the galaxy through the cluster. In a second test we assume that the nucleus has moved from the edge of the inner lobe, $\sim$9~kpc, which should give an upper limit to the proper motion. 
%We thus assume that the inner parts of the low frequency lobes were formed since the start of the current epoch of jet activity.
The original location of the nucleus is unknown, but the distance travelled by the real nucleus is likely bounded by both numbers. 

Assuming any reasonable travelled distance allows us to calculate a proper motion and determine if this proper motion is reasonable considering the known radial velocities in this cluster of galaxies. Taking a hotspot advance speed of 0.02c \citep{muxlow88} gives a jet activity age of 2$\times$10$^7$ years.
%If we would assume that the low frequency lobe is part of the relic lobe,then the derived proper motion will be much smaller. 
We thus obtain a proper motion of 0.24~$\mu$as yr$^{-1}$ or $\sim$250~km~s$^{-1}$ if the nucleus has moved from the midpoint between both lobes. Assuming the nucleus moved the full distance between both lobes, the proper motion is 0.45~$\mu$as yr$^{-1}$ or $\sim$450~km~s$^{-1}$. Taking the slower hotspot advance speed of 0.005c \citep{alexander96} the above velocities decrease by a factor of 4. Assuming that the low-frequency plasma is only half the age of the current phase of jet activity (following \citealt{blundell00}) doubles the values. But even the maximum value of 900~km~s$^{-1}$ is not an unreasonably large proper motion in this cluster of galaxies. 
%This is of the same order, although a bit smaller, as the sound speed derived for a plasma with a temperature of 2~keV. 

\cite{ledlow05} measured a radial velocity offset of 2197~km~s$^{-1}$ towards the Earth for Cygnus A, compared to the mean cluster radial velocity. However, those authors suggest that the cluster could be composed of 2 sub-clusters. If so, then for the model also taking positional data into account, Cygnus A has a radial velocity offset from the closest subcluster of $\sim$ 1000~km~s$^{-1}$. However this result is rather dependent on the modelling and assumed sub-cluster membership and could be as small as 163~km~s$^{-1}$ \citep{ledlow05}.  The sub-cluster Cygnus~A likely belongs to has a radial velocity dispersion of 1134~km~s$^{-1}$ \citep{ledlow05}. Thus over 2$\times$10$^7$ years the nucleus could have moved along the line-of-sight anywhere between 3.3 to 20.4~kpc compared to the subcluster movement. It is thus not unreasonable that the galaxy also has a proper motion and has travelled 5$-$9~kpc in the plane of the sky. 

The radial velocity offset measured for Cygnus~A is thus of the same order of the proper motion derived above. Both measurements also have very large uncertainties. The above radial velocities indicate that the proper motion we calculate is within the expected range. We conclude that proper motion, is the best explanation for the superposition of the nucleus with the counterlobe. The projection effect would be due to the radial velocity of the nucleus toward us with respect to the rest of the sub-cluster and cluster. From the morphology of the inner counterlobe, which becomes wider and shows a ``southern spur'' (see Fig.~\ref{fig:nucleus2}), the counterlobe plasma seems partially deflected, potentially by the galaxy, which would indicate that the nucleus is located physically in the counterlobe. But there are other ICM perturbations that bend the lobe and counterlobe plasma, and thus morphology alone cannot demonstrate whether this is just a superposition or not.

% A possible method of testing whether the nucleus is located in front or in the counterlobe, is to study high-resolution low frequency polarised images of the nucleus and its immediate surroundings. The nuclear emission is normally unpolarised, and would appear so if located in front of the counterlobe. However, if the nucleus is located within (or behind) the counterlobe, we should observe polarised low-frequency emission due to the lobe plasma in front of the nucleus. Thus low-frequency polarised emission would be evidence for the nucleus being located in the counterlobe. 

Could this proper motion be detectable with VLBI? \cite{krichbaum98} and \cite{bach04} have studied VLBI images taken over 2 and 7 years apart, respectively. However, \cite{bach04} does not identify the nucleus with any certainty, and \cite{krichbaum98} aligned their images at different frequencies so that their assumed nucleus component is stationary. Hence no proper motions or upper limits are obtained by either set of authors. The expected distance travelled for an assumed proper motion of 900~km~s$^{-1}$, which is the largest proper motion expected, in the 7 years of VLBI studies, is $\sim$6$\times$ 10$^{-6}$ arcsec, and thus below the detection limit of the VLBI. The current spatial resolution of the Global Millimeter VLBI Array (GMVA) is 40$-$50~$\mu$as, with a likely improvement in the near future to 15$-$20~$\mu$as \citep{krichbaum06}. Thus direct measurement of any displacement of the nucleus of Cygnus~A might become possible in the not too distant future.

%Possible other pieces of evidence for the proper motion of the nucleus are an asymmetry in the jet speed and potentially the direction. The largest proper motion calculated above of 900~km~s$^{-1}$, or 0.003c, is much smaller (by a factor of $\sim$115) than the best fit jet speed of $\sim$0.35~c \citep{steenbrugge08a}. We thus do not expect to detect the nuclear velocity shift from the precession modelling of the counterjet versus jet, certainly as the counterjet is much weaker and therefore difficult to trace even at 15~GHz.

\subsubsection{Comparison with wide-angle tailed radio galaxies}

Radial velocity and proper motion through a cluster due to a cluster merger is the generally accepted explanation for the wide angle tailed (WAT) radio galaxies \citep{sakelliou00}. Like Cygnus~A, WAT radio galaxies are offset from the cluster's centre determined from X-ray images. Why then do we hardly see any morphological evidence for Cygnus~A's movement through its cluster or sub-cluster?  The bending of the lobes observed in WAT radio galaxies depends on the ram pressure exerted on the lobes by the ICM. WAT radio galaxies are located in richness class 0$-$3 clusters, and are thus not necessarily located in only the densest clusters \citep{sakelliou00}. However, according to \cite{mao10} WAT sources are only detected in the densest regions of the cluster. There are some indications that Cygnus~A is located in a low density part of the cluster. Namely, FR~II galaxies are generally observed in poor groups, where densities are lower; and FR~I radio galaxies are normally observed at the centres of clusters of galaxies. Cygnus~A is one of only a handful of FR~II galaxies located in a cluster of galaxies. Thus a difference in density at the location of Cygnus~A in comparison with WAT radio galaxies in their respective clusters could be an explanation for the difference in morphology. Space velocities of WAT radio sources, as for Cygnus~A, are generally poorly known, but a few 100~km~s$^{-1}$ seem to be sufficient to create WAT radio galaxies \citep{sakelliou96}. Thus a difference in space velocity is an unlikely explanation for the difference in morphology. The rather young age of the Cygnus~A lobes might be a further explanation for the small effect the motion through the cluster has on the morphology of the lobes. 
Although the cluster plasma density at the location of Cygnus~A is lower than in the center of other clusters, at 15~GHz there are indications that the lobe is confined by the intracluster medium. This is consisitent with the conclusion from \cite{barthel96} that the slower adiabatic expansion of the radio lobes due to the larger density of the cluster environment compared to the field increases the radio luminosity.

We should note that any proper motion of the nucleus with respect to the lobes would violate the assumptions made by \cite{steenbrugge08a} in their section 4.5. In that section the jet speed was determined from the distance from the nucleus to pairs of jetknots observed in the jet and counterjet. The jet knots closer to the nucleus were emitted quite recently compared to the start of the jet activity phase, and the derived speed is thus less affected. It also invalidates for this source the method, by which the jet speed is determined from the jet/counterjet lengths, if these lengths are determined as the distance between the nucleus and the hotspots \citep{alexander96}. A proper motion of the nucleus also affects the estimate for the duration of the previous epoch of jet activity, as this was determined from the length of the relic counterjet, which stops just short of the current hotspots, and the current position of the nucleus. 
%This estimate depends on the assumption that the hotspot advance speed was the same during both epochs of jet activity, and is thus rather uncertain. the proper motion of the nucleus would make this value a lower limit for a hotspot advance speed that is smaller or equal to the current value.  

%One explanation for the current jet precession explored by \cite{steenbrugge08a} is the presence of a possible second nucleus detected by \cite{canalizo03} using Keck IR images. However, the current distance between the bright X-ray nucleus and the possible second nucleus is rather large, 400~pc, or 0.38$^{\prime\prime}$, in projection southwest of the X-ray bright nucleus. With any proper motion for one or both of the nuclei, both nuclei could have been much closer in the past, depending on the direction of the proper motion(s). Hence this makes the possible second nucleus a more likely explanation for the observed jet precession.

\subsubsection{Jet-counterjet angle}

In \cite{steenbrugge08a} we noted that the best precession model does require an angle of about 179$^{\circ}$, rather than the expected 180$^{\circ}$ between the inner jet and counterjet. In Fig.~\ref{fig:nucleus3} we show the inner jet and counterjet, with a dashed line through the inner jet which we extend on the counterlobe side and vice-versa. The weak counterjet lies below this dashed line, indicating that the jet and counterjet do not make a 180$^{\circ}$ angle. The same figure, but without the lines tracing the jets, is shown in Fig.~\ref{fig:nucleus4}. 

The proper motion of Cygnus~A's host galaxy can explain the observed angle difference. If we assume that the nucleus has a displacement of 2~kpc from the trajectory of the counterjet due to its proper motion, then using the lengths of the jet and counterjet, 140.6~kpc and 122.3~kpc respectively, we can calculate the angle between the jet and counterjet, and thus the difference from 180$^{\circ}$. This difference in angle is less than 1 degree, namely 0.87 degrees for the values given above, providing a plausible explanation for the observed jet-counterjet angle. However, we should note that there is a substantial uncertainty in the calculated value. There are uncertainties on the angle of the jet with our line of sight, and the displacement from the original position is unknown. We conclude that the proper motion of the nucleus can explain why the observed angle between the inner jets and the nucleus is 179$^{\circ}$, instead of the predicted 180$^{\circ}$. In this scenario the jets are emitted at a 180$^{\circ}$ angle, and only due to the proper motion of the nucleus compared to the jets appears as a 179$^{\circ}$. For a more accurate calculation deeper, high-resolution radio data are needed.

\section{Conclusion}
In this article we studied the position of the nucleus with respect to the lobes as well as the trajectory of the inner jets. We found that the nucleus is located in front of or in the counterlobe, the best explanation for which is a proper motion of the Cygnus~A galaxy through the Cygnus cluster, probably caused by the ongoing merger in this cluster. The derived range in proper motion, between 60 and 900~km~s$^{-1}$ has the same order of magnitude as the range in radial velocity offsets determined by \cite{ledlow05}. The lack of bending over larger angles as seen in WAT radio sources seems to indicate that Cygnus~A is located in a rather low density region compared to central cluster densities. This proper motion is also likely to be the explanation that the observed angle between the inner jet and counterjet with the nucleus is 179$^{\circ}$ as determined from the precession modelling. Future, deep and high-resolution low-frequency polarisation images should be able to determine whether the nucleus is located in or in front of the counterlobe.

%The proper motion of the nucleus and galaxy in the direction toward the counterlobe could explain the higher in density on the counterlobe side.

\begin{acknowledgements}

KCS and SP gratefully acknowledge the support from ALMA-CONICYT grant 31110019, and to ESO for its flexibility so that ESO project money could be used till CONICYT transferred the grant several months after the project started.
\end{acknowledgements}

\bibliography{references}

\end{document}